\documentstyle[12pt,psfig]{article}

\makeatletter

\def\section{\@startsection {section}{1}{\z@}{24pt plus 2pt minus 2pt}
{12pt plus 2pt minus 2pt}{\large\bf}}

\def\subsection{\@startsection {subsection}{2}{\z@}{12pt plus 2pt minus 2pt}
{12pt plus 2pt minus 2pt}{\normalsize\bf}}
\makeatother

\begin{document}

\date{}


\title{\Large\bf PROGRAPE-1: A Programmable, Multi-Purpose Computer for Many-Body Simulations}

\author{Tsuyoshi Hamada\thanks{Department of General Systems Studies, 
College of Arts and Sciences, University of Tokyo, 
Tokyo 153, Email: hamada@grape.c.u-tokyo.ac.jp},  
Toshiyuki Fukushige\thanks{Department of General Systems Studies, 
College of Arts and Sciences, University of Tokyo, 
Tokyo 153, Email: fukushig@provence.c.u-tokyo.ac.jp}  
Atsushi Kawai\thanks{Department of General Systems Studies, 
College of Arts and Sciences, University of Tokyo, 
Tokyo 153, Email: kawai@grape.c.u-tokyo.ac.jp}  
\\ and \\ 
Junichiro Makino\thanks{Department of Astronomy, 
School of Science, University of Tokyo, 
Tokyo 113, Email: makino@astron.s.u-tokyo.ac.jp}\\ 
University of Tokyo
}

\maketitle

\subsection*{\centering Abstract} 
 We have developed PROGRAPE-1 (PROgrammable GRAPE-1), a
programmable multi-purpose computer for many-body simulations. The
main difference between PROGRAPE-1 and ``traditional'' GRAPE systems
is that the former uses FPGA (Field Programmable Gate Array) chips as
the processing elements, while the latter rely on the hardwired
pipeline processor specialized to gravitational interactions.  Since
the logic implemented in FPGA chips can be reconfigured, we can use
PROGRAPE-1 to calculate not only gravitational interactions but also
other forms of interactions such as van der Waals force,
hydrodynamical interactions in SPH calculation and so on.  PROGRAPE-1
comprises two Altera EPF10K100 FPGA chips, each of which contains
nominally 100,000 gates. To evaluate the programmability and
performance of PROGRAPE-1, we implemented a pipeline  for
gravitational interaction similar to that of GRAPE-3. One pipeline
fitted into a single FPGA chip, which operated at 16 MHz clock. Thus,
for gravitational interaction, PROGRAPE-1 provided the speed of 0.96
Gflops-equivalent. PROGRAPE will prove to be useful for wide-range of
particle-based simulations in which the calculation cost of
interactions other than gravity is high, such as the evaluation of SPH
interactions.

\section{Introduction}

GRAPE ("GRAvity piPE"; \cite{scmieu90}\cite{mt98}) is a
special-purpose computer for the calculation of the gravity in the
astronomical many-body simulations. It has hardware specialized for
the calculation of the gravity, which is the most expensive parts in
the astronomical many-body simulations. Its only function is to
calculate the gravitational interactions between particles.  All other
calculations, such as I/O, time integration and diagnostics, are
handled by a general-purpose computer (the ``host computer'')
connected to GRAPE.  This hybrid architecture has proved itself quite
useful, because the special-purpose nature of GRAPE hardware made it
possible to achieve very high performance with relatively low cost,
while the programmability of the host computer made it possible to
apply the same GRAPE hardware to wide variety of astrophysical
problems, from formation of Moon \cite{kim99} to the formation of
galaxies using $N$-body approach \cite{fm97} or SPH approach ({\it
e.g.} \cite{ns97}\cite{myn99}. See Makino and Taiji (1998) \cite{mt98}
or Hut and Makino (1999) \cite{hm99} for reviews.

For simulations of pure $N$-body systems, GRAPE hardwares offer huge
advantage both in the absolute speed and in the
price-performance. However, though many astrophysical phenomena are
primarily driven by gravitational force, hydrodynamics and other
physics are sometimes important.  A number of research groups use
GRAPE hardwares mainly for combined $N$-body+hydrodynamical
simulations of formation and evolution of galaxies or clusters of
galaxies ( \cite{ufmestl93}\cite{s96}\cite{k97}\cite{nmn97}). All
implementations of SPH on GRAPE use it only to calculate gravitational
interaction and to construct the list of neighbors for SPH
interactions. Actual evaluation of SPH interaction is performed on the
host computer.

Thus, in SPH calculations on GRAPE, the speed of the host computer
tends to determine the total performance of the system, since the
calculation cost of the SPH interaction is, though not as large as
that of the gravitational interaction, much larger than the
calculation cost of the rest of the simulation program such as time
integration and I/O.

The calculation cost of the SPH interaction is rather high. The
average number of neighbors of a particle in modern SPH programs is
around 30 to 60, and calculation cost of single SPH interaction is a
few times more than that of gravitational interaction (since the
expression is more complex). Thus, when one uses the tree algorithm
\cite{bh86}, the calculation cost of SPH interaction is not
much smaller than that of gravity.  If we perform a pure SPH
simulation without collisionless particles, the gain in speed achieved
by GRAPE is a factor of few at the best. In practice, in many SPH
simulations relatively large number of collisionless particles are
included, and for such calculations GRAPE offers a large speedup.
However, when we want to perform SPH simulations without collisionless 
particles, the speedup would be rather limited. Accordingly, though SPH
techniques have been applied to wide variety of problems, such as the
dynamics of star-forming regions and hydrodynamical interaction of
stars, GRAPE  hardwares have not been widely used for those kind of
simulations.

In principle, one could develop a special-purpose hardware similar to
GRAPE to accelerate SPH interactions \cite{yoicm99}. Such a
hardware, once completed, would offer a large speedup over
general-purpose computers. 

However, so far the viability of such a
project has remained unproven. There are several reasons why it is
difficult to develop a specialized computer for SPH. First, the
calculation of the SPH interaction is quite a bit more complex
compared to the calculation of the gravitational interaction. With
gravity, the only thing we need to implement is the force which is
proportional to the inverse square of the distance. The pipelined
circuit to evaluate this interaction comprises around 15 arithmetic
units (adders and multipliers) and one unit to evaluate $f(x) =
x^{-3/2}$. In concept, the hardware for SPH is equally simple. What we need is a
hardware to evaluate sum of quantities weighted by the kernel function
$W(r;h)$ or its derivative.  In practice, there are numerous technical
details which have to be taken care of in hardware. For example, the
smoothing length $h$ is different for different particles, and the
kernel $W$ has to be symmetrized. 

The second reason is that there are
rather large number of varieties in SPH algorithms.  Several different
ways to symmetrize $W$ are used. In addition, there are many methods
to implement artificial viscosity.  For some problems, some methods
work fine. For some other problems, some other methods seem to be
better. Moreover, there have been several new developments in the last
few years. If we design a hardware for a specific SPH algorithm,
the development the hardware would be significantly more difficult and
time-consuming compared to the development of a GRAPE, and yet the
hardware might become obsolete even before its completion.

We can avoid the risk of becoming obsolete if we can change the
hardware of the pipelined processor for SPH {\it after its
completion}. Changing the hardware might sound self-contradictory,
since the hardware is, unlike the software, cannot be changed once it
is completed. Though the name might sound strange, such
``programmable'' chips have been available for several decades
now. These chips, usually called FPGA (Field-programmable gate array)
chips, consist of small logic elements (LE) and switching matrix to
connect them. A LE is typically a small lookup table made of an SRAM,
combined with additional circuits such as a flip-flop and special
logic for arithmetic operations. The design procedure for an FPGA is
largely similar to that for traditional gate arrays; one writes the
design either in the schematics or in hardware description languages
such as VHDL, and then the CAD software assigns the logic to LEs and
generates the connection pattern for the switching matrix. The design
is loaded to the FPGA chip either from a small ROM chip or from a
dedicated write port.

The size (in equivalent gate count) of FPGA chips has been enormously
increased, from around 1K of mid-1980s to around 100K of
1997. Since the increase is driven essentially by the advance in the
semiconductor device technology, we can expect that the increase will
continue for the next several years. An FPGA chip with over one
million gates will be available by 2001. Of course, one should 
not take these advertised numbers for its face value. Actual 
size of the circuit which can be implemented on a particular FPGA chip 
is limited by various factors,  much in the same way as a
program rarely runs at the theoretical peak speed of the computer used. 

The size of the present (and future) FPGA chip is large enough to
house fairly complex circuits. For example, the pipeline processor
chip of GRAPE-3 \cite{omefis93} is around 20K gates, and
that of GRAPE-4 is around 100K.  Thus, even if we assume rather low
utilization ratio of 20\%, a GRAPE-3 chip should fit into an FPGA
chip with 100K gates.

Of course, to implement a GRAPE on FPGA would have little practical
meaning, since GRAPE pipelines implemented on custom LSI
chips of similar price to FPGAs are much faster. However, applications
for which no custom hardware is available would find a GRAPE-like
machine implemented using FPGA useful. Consider the case of SPH.  We
can implement any variety of SPH algorithm, as far as it is expressed
as interaction between particles and the necessary circuit fits in a
target FPGA chip. The peak performance is not as large as what we can
enjoy with custom LSI chips, but is still orders of magnitude faster
than what is available on general-purpose computers of similar price.

We call the concept of using a programmable FPGA chip as the pipeline
processor as PROGRAPE (PROgrammable GRAPE). It should be noted that
PROGRAPE is not the first attempt to use FPGAs as the building blocks
of special-purpose computers.  There are a number of projects to
develop custom computing machines using FPGAs as a main unit. The most
influential are probably Splash-1 and Splash-2 project \cite{bak96}.

Projects like Splash did not select a specific application area as
their target. Thus, machines developed in these projects can be
regarded as ``general-purpose''.  In the sense, one could use them for
any application. One of the unique features of PROGRAPE is that its
architecture is specialized to a rather limited range of problems,
namely the evaluation of the particle-particle interaction in
many-body simulations. Thus, there is relatively little room of
programmability. The only thing one can change is the functional form
of interaction between particles, and all else are essentially
fixed. This might sound like a limitation, but it actually means the
amount of work of both the designer of the PROGRAPE hardware and the
application programmer using it can be greatly reduced, compared to
other FPGA-based machines like Splash.  With Splash everything on
board is programmable, and that means the user must program everything
on board. If our main interest is in particle-based simulation, we do
not need the programmability other than the possibility to change the
interaction, and PROGRAPE is designed to provide just the minimal
amount of programmability needed to be useful.

We have developed a GRAPE-like hardware based on FPGA chips which we
call PROGRAPE-1 (PROgrammable GRAPE-1). It contains two Altera
EPF10K100 FPGA chips (around 100K nominal gates).  We tested its
performance with a realistic application by implementing the pipeline
similar to that of GRAPE-3. One pipeline was successfully fitted into
one chip, and operated at 16 MHz clock. Thus, the performance of the
single FPGA chips was shown to be comparable to that of a GRAPE chip
used in GRAPE-3. We are currently working to implement an SPH
calculation code.

In section 2, we describe the concept of the PROGRAPE system.  In
section 3, we describe the hardware design of PROGRAPE-1.  In section
4, we describe the software and example application of
PROGRAPE-1. Section 5 is for discussions.

\section{PROGRAPE System}

In this section, we describe the basic idea behind the PROGRAPE
approach to build multi-purpose programmable computer for
particle-based simulations. In section 2.1., we describe the hardware
architecture, and in 2.2, how we use the hardware.

\subsection{Hardware architecture}

PROGRAPE is a programmable multi-purpose computer for many-body
simulations. Figure~1 shows the basic structure of a PROGRAPE system.
The system consists of a PROGRAPE and a host computer. The PROGRAPE
calculates interactions between particles. The host computer performs
all other calculations.

\begin{figure}
\centerline{\psfig{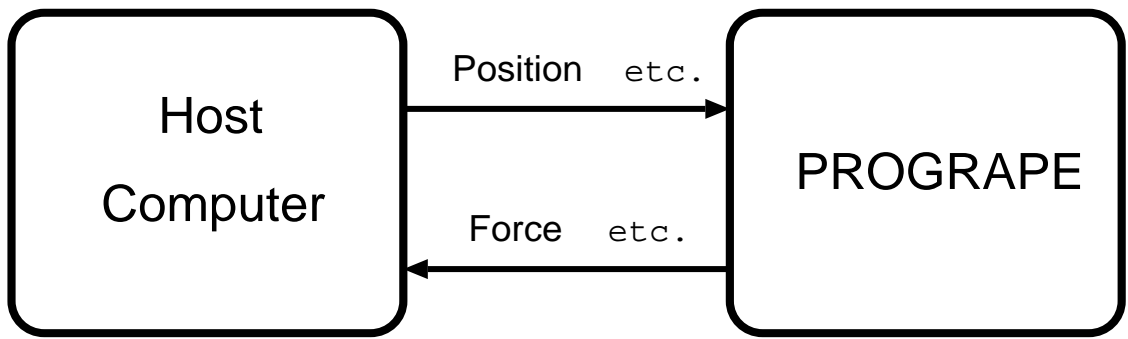}}
\caption{Basic structure of the PROGRAPE system.}
\end{figure}

\def\bm#1{\mbox{\boldmath $#1$}}

Figure~2 shows the structure of a PROGRAPE. It consists of a control
unit, an interface unit, a memory unit and multiple FPGA chips. This
structure is the same as that of GRAPE hardwares such as GRAPE-3,
GRAPE-4 and GRAPE-5.

\begin{figure}
\centerline{\psfig{file=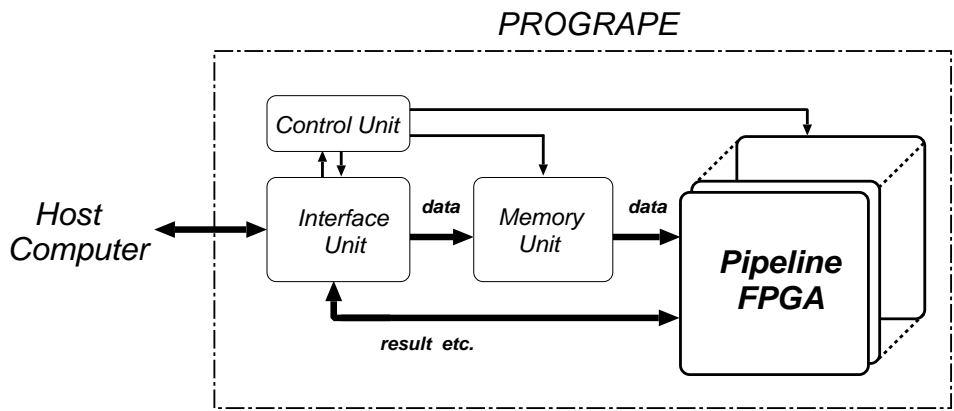,width=150mm}}
\caption{The hardware architecture of PROGRAPE.}
\end{figure}

In a rather abstract expression, a PROGRAPE evaluates the following:
\begin{equation}
\bm{f}_{i} = \sum_{j=1}^{N}\bm{g}(\bm{X}_{i},\bm{X}_{j}),
\end{equation}
where $\bm{X}_{i}$ is the vector which contains the 
information of particle $i$, $\bm{f}_i$ is the resulted ``force'' on
particle $i$, and $\bm{g}$ is a function which describes the
interaction between two particles.

To give a specific example, in the case of GRAPE for gravitational
force, $\bm{X}_i = (\bm{x}_i, m_i)$ where  $\bm{x}_i$ and  $m_i$ are
the position and mass of particle $i$, and $\bm{g}$ would be given by

\begin{equation}
\bm{g}= -m_j\frac{\bm{x}_i - \bm{x}_j}
                    {(|\bm{x}_i - \bm{x}_j|^2 + \epsilon^2)^{3/2}},
\end{equation}
where $\epsilon$ is the softening parameter. 
The result, $\bm{f}_i$, is the gravitational acceleration of particle
$i$. Note that $\bm{X}_i$ and  $\bm{X}_j$ in equation
(1) do not necessarily  contain the same amount of
information, though we refer to them by the same symbol for
simplicity. In the case of the gravitational interaction, mass $m_i$
is not used to calculate the acceleration $\bm{f}_i$. Therefore, we
can remove $m_i$ from $\bm{X}_i$, though $\bm{X}_j$ should contain
$m_j$. If  a PROGRAPE has multiple pipeline FPGA chips (or if an FPGA
chip  houses multiple pipelines), $\bm{f}_i$ for multiple particles
are evaluated in parallel.

If PROGRAPE is used to evaluate the SPH interaction, $\bm{X}$ will contain 
density and pressure to calculate the pressure gradient, and velocity
to calculate the artificial viscosity. When variable smoothing is
used, the size of the smoothing length must also be included in
$\bm{X}$. The interaction function $\bm{g}$ will be more complex than
equation (2).

The calculation proceeds in the following steps. First, the host
computer sends the configuration data of pipeline FPGA chips. This
need to be done only once. Then, the host computer stores the data of
particles into the memory unit. These particles are used as $\bm{X}_j$
in equation (1), and we refer to them as
``$j$-particles''.  After sending the data of $j$-particles, the host
computer sends  $\bm{X}_i$ to each pipeline.  These data will be
stored in  registers in the pipelines, which we call $i$-registers
($i$ denotes $i$-particles, as opposed to $j$-particles). Next, the
host computer sends the command to the control unit to start the
calculation. Accordingly, the control units sends address (particle
index) to the memory unit and control signals to the pipeline FPGA
chips. Pipeline FPGA chips receive the data from the memory, evaluate
the interaction function $\bm{g}$ and accumulate them in their
internal accumulators. Finally, when the summation is finished, the
contents of the accumulators are sent back to the host computer.

Figure 3 shows the conceptual design of the circuit in a pipeline
FPGA chip. As stated above, a pipeline FPGA chip consists of a memory
interface, an I/O interface and a pipeline unit. A pipeline unit
consists of an arbitrary number of pipelines including one. A
pipeline consists of $i$-registers, an interaction unit to evaluate
an interaction function, and accumulators.

\begin{figure}
\centerline{\psfig{file=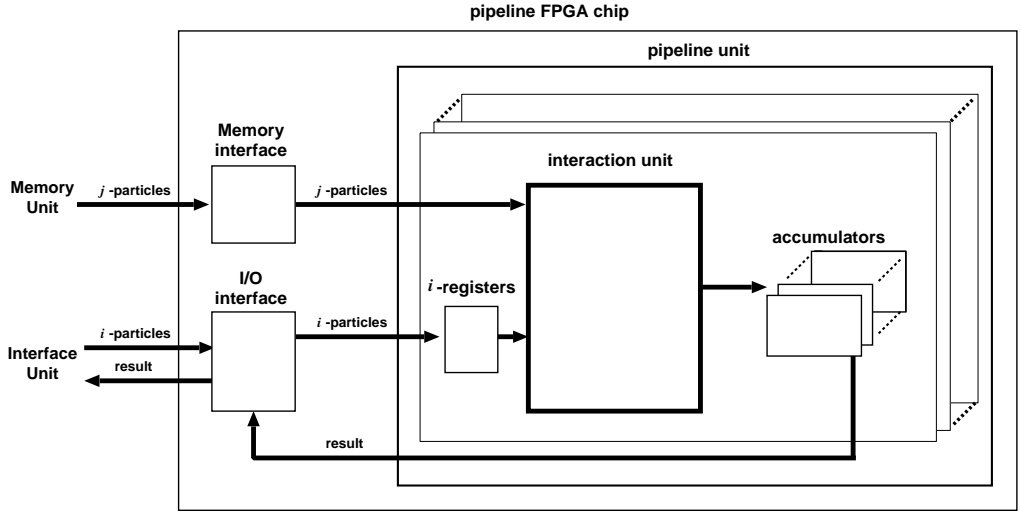,width=150mm}}
\caption{The conceptual design of the circuit in a pipeline chip of PROGRAPE.}
\end{figure}

Note that we can change only the content of the pipeline FPGA
chips. All other parts of hardware such as an interface unit, a
control unit and a memory unit are fixed. For example, once the
hardware is designed and built, the amount of data transferred from
the memory unit to the pipeline FPGA chips in one clock cycle is
fixed. However, this is not a serious limitation, since even if
$\bm{X}_j$ requires the number of bits larger than what is available
in hardware, we can still use the PROGRAPE hardware by designing the
pipeline FPGA chip so that it uses more than one clock cycles to
receive $\bm{X}_j$. In the case of PROGRAPE-1 described in section 3,
the memory unit can transfer 16 bytes of data per clock cycle. For
most applications, this is more than necessary, and we do not have to
design a complex multi-cycle pipeline.

\subsection{Software architecture}

Once the content of the pipeline is designed and tested, the way we
use a PROGRAPE system is very much the same as the way we use any of
GRAPE systems. The application program sends the data of $j$-particles 
and $i$-particles, and the PROGRAPE hardware returns the result
calculated  for $i$-particles. From the point of view of the
application program, only a small number of library functions are
visible.

The big difference between PROGRAPE and GRAPE is that in the case of
PROGRAPE, the user {\it can} specify every detail of the pipeline. To
change its logic in a slightly different way, the user {\it must}
specify literally every bit of the pipeline.

It should be noted that it is also true that the user need to specify
only pipeline FPGA chips. As shown in figure 2, a PROGRAPE hardware
consists of a control unit, an interface unit, a memory unit and
pipeline FPGA chips. Everything other than pipeline FPGA chips are
fixed. Also, a memory interface , an I/O interface, $i$-registers and
accumulators in pipeline FPGA chips are pre-designed and provided as
parametrized modules.

Figure~4 shows the conceptual design of a pipeline in a pipeline unit
in the VHDL language. A pipeline receives data from an I/O interface
and a memory interface and evaluates an interaction. In figure~4, the
{\tt Xi} and {\tt Xj} are the ports of data put from an I/O interface
and a memory interface respectively. The evaluated result is then fed
to the accumulator module. Compared to programming other
FPGA-based machines, programming a PROGRAPE is much easier, simply
because the large part of the hardware is already programmed.

\begin{figure}
\begin{verbatim}
entity pipeline is 
    generic ( 
        JDATA_WIDTH : integer;
        IDATA_WIDTH : integer;
        FDATA_WIDTH : integer;
        ADR_WIDTH   : integer;
        NVP : integer
    );
    port( 
        Xj : in std_logic_vector(JDATA_WIDTH-1 downto 0);
        Xi : inout std_logic_vector(IDATA_WIDTH-1 downto 0);
        adr   : in std_logic_vector(ADR_WIDTH-1 downto 0);
        we, run, iclk, pclk : in std_logic;
        RESULT : out std_logic_vector(FDATA_WIDTH - 1 downto 0)
    );
end pipeline;

architecture std of pipeline is

signal idata : std_logic_vector(IDATA_WIDTH-1 downto 0);
signal fdata : std_logic_vector(FDATA_WIDTH - 1 downto 0);
signal runr  : std_logic;

begin

    u1: i_register GENERIC MAP(IDATA_WIDTH => IDATA_WIDTH, NVP => NVP)
           PORT MAP(datai => Xi, we =>we, adr =>adr, run => run,
              idata => idata, iclk => iclk, pclk => pclk);

    u2: interaction GENERIC MAP(JDATA_WIDTH => JDATA_WIDTH,
              IDATA_WIDTH => IDATA_WIDTH)
           PORT MAP(jdata => Xj, idata => idata, run => run,
              runr => runr, fdata => fdata, clk => pclk);

    u3: accumulators GENERIC MAP(FDATA_WIDTH => FDATA_WIDTH, NVP => NVP)
           PORT MAP(fdata => fdata, run => runr, adr => adr, 
              datao => RESULT, iclk => iclk, pclk => pclk);

end pipeline;
\end{verbatim}
\caption{The conceptual design of a pipeline of PRPGRAPE in the VHDL language. }
\end{figure}

Even though programming PROGRAPE is not as hard as programming other
FPGA-based machines, the design and testing of a pipelined FPGA chip
for particle-particle interaction is, judging from our experience of
developing custom LSI chips for GRAPE hardwares, still a rather
time-consuming work. As we stated earlier, the design process of
the internal logic of an FPGA is essentially the same as that for a
custom LSI, except that the risk of making mistakes is small.

In practice, this smaller risk is a  significant factor. With a
custom LSI, a serious mistake in the design would cost several tens of
million yens and cause the delay of several months in schedule. 
Even if one makes no mistake, several tens of million yens is still
necessary to get first samples of the chip.  On the other hand, even
most advanced FPGA chips would cost around hundred thousand yens. In
addition, if the mistake is found in the design of an FPGA, it can be
fixed in a few hours or even minutes. This difference means the
iteration in the design is much quicker with FPGA, and therefore the
design process as a whole is also a lot quicker. Even so, the design
of the pipeline still looks like a serious project, which would
require considerable investment even for an expert in programming
general-purpose computers.

There are several reasons for the difficulty of the programming. The
first obstacle is that there are too much freedom. An FPGA chip can be
used to implement any logic circuit, as far as it fits into the chip.
Thus, the programming of FPGA is done essentially in the lowest level of
logic, in which one specifies basic logic gates (AND/OR gates) and
flip-flops. To make an analogy, an FPGA is something like a universal
Turing machine without any high-level language, library functions or
even operating systems. Thus, when one designs a machine using FPGA,
all these infrastructure must also be developed by someone.

This complete freedom is available for all operations to be
implemented in FPGA. Thus, even for a simple addition, there are
infinitely many possibilities, which would cause very large variation
in the size and speed of the implemented circuit. For many
applications, it would be necessary to determine what number format
and accuracy are appropriate for each operation, otherwise, the
pipeline would not fit into available FPGA chips. Even if they fit,
the performance would be rather low. Thus, unlike a program on a
general-purpose computer, a design of the pipeline for a PROGRAPE
requires in-depth understanding of both the arithmetic operations in
digital circuit and the effect of the round-off error in the target
application. In many application areas, little is known about the
effect of the round-off error. Thus, in many case the user need to
conduct systematic experiments to understand the error propagation
mechanism.

In principle, if FPGA chips are large and fast enough, we can just
implement standard 32- or 64-bit floating-point arithmetic operations. 
Unfortunately, we need to wait several years before such large FPGA chips
become available. Currently, even the largest FPGA chips would not
be able to house a single 64-bit multiplier.  Moreover, even when large
chips are available, we can achieve very large speedup by adjusting
the size of arithmetic units according to the required accuracy. Thus,
to achieve a good performance, it will always be necessary to select right
arithmetic units for each operation.

The second obstacle is the simple fact that what is designed is a
hardware, not a software. Though the design of a hardware is not
necessarily more difficult than that of the software, they are
different and require different kinds of expertise. To some extent,
the situation is similar to that of programming parallel
computers. The programming of a parallel computer is not necessarily
more difficult than that for a sequential computer. However, the lack
of experience and the lack of good software environment make the
development of parallel program a difficult task. This is true of the
case of the hardware design.

At present, those who wish to use PROGRAPE must write the design of
the pipeline in VHDL, unless the application has been developed by
somebody else. We are  developing some kind of 
design environment  which will be discussed elsewhere.

\section{PROGRAPE-1}

PROGRAPE-1 is the first machine of the PROGRAPE architecture. Figure~5
shows the block diagram of PROGRAPE-1.  It is connected to the host
computer through a PCI interface board \cite{kftms97}. The interface
protocol used between PHIB and PROGRAPE-1 is the Hlink protocol
described in Makino {\it et al.} \cite{mtes97}.

\begin{figure}
\centerline{\psfig{file=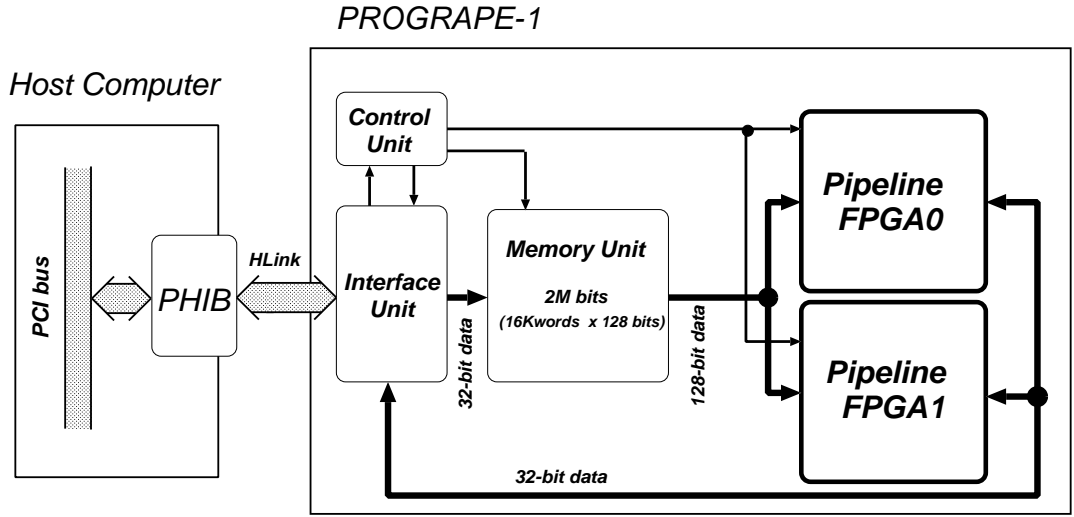,width=150mm}}
\caption{Block diagram of the PROGRAPE-1 system.}
\end{figure}

The PROGRAPE-1 board consists of two pipeline FPGAs, a memory unit, an
interface unit and a control unit. Thus, PROGRAPE-1 is a minimal
realization of the concept of PROGRAPE (figure 2) with two pipeline
FPGA chips.  In the following subsections, we will describe each of
these units.

\subsection{Pipeline FPGA}

We adopted Altera EPF10K100 chips as the pipeline FPGA.  This FPGA has
4992 logic cells, each of which has a look-up table with 4-bit input
and a flip-flop, and 12 SRAM blocks, with the size of 2 Kbits each.
The manufacturer claims that the chip has the capacity equivalent to
about 100,000 gates.

This FPGA chip from Altera was not the only possibility. Similar chips
were available from Xilinx and Lucent Technologies. The main reason we
chose Altera was that we had some experience with its design software.

The configuration data of the pipeline FPGAs are written from a
special port connected to the interface unit. Thus, the pipeline can
be programmed by software on the host computer.

Figure~6 shows the I/O specification of the pipeline FPGA. 
It has one data input port ({\tt JDATA[127:0]}), one bidirectional
data port ({\tt IDATA[31:0]}), one address port ({\tt ADR[9:0]}), four
control input pins ({\tt CS, RE, WE, RUN}) and one clock pin ({\tt
CLK}).  The data input port ({\tt JDATA}) is used to supply the data
of $j$-particles from the memory unit.  The bidirectional data port
({\tt IDATA}) and address port ({\tt ADR}) are used to read or write
the on-chip registers, whose organization is specified by the
configuration data fed from the host computer.  The control pins ({\tt
CS}, {\tt RE}, {\tt WE)} are used to control the read and write
operations to the on-chip register.  The {\tt CLK} pin supplies the
clock signal.  The {\tt RUN} pin is used to start operations of a
interaction unit and accumulators in a pipeline unit of a pipeline
FPGA chip.

\begin{figure}
\centerline{\psfig{file=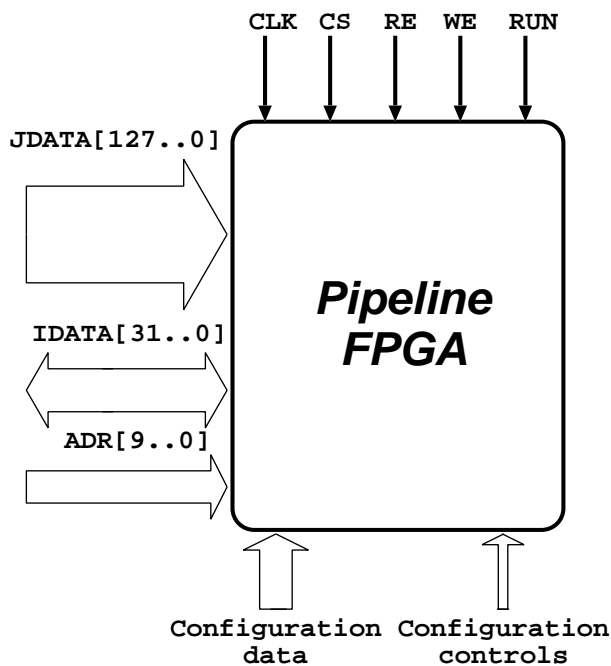,width=100mm}}
\caption{I/O specification of the pipeline FPGA.}
\end{figure}

\subsection{Memory Unit}

The memory unit supplies the data of $j$-particles to pipeline FPGA
chips according to the address supplied by the control unit. Two
pipeline FPGA chips receive the same data.  The memory unit consists
of four 512 Kbit (16 Kwords $\times$ 32bits) SRAM modules. The width
of the data port from the memory unit to pipelines is 128-bit
width. The width of the data port from the interface unit to memory
unit is 32-bit width.  We adopted 7MC4032 made by Integrated Device
Technology, Inc, which is 512K bits SRAM module with separate I/O.

\subsection{Interface Unit}

The interface unit handles the communication between the host computer
and PROGRAPE-1. There are 5 communication modes: (1) The host computer sends a
command to the control unit. (2) The host computer sends the data to memory
unit. (3) The host computer sends the data to the pipeline FPGA chips. (4) The
pipeline FPGA chips send the calculated result to the host computer. (5) The
host computer sends the configuration data to pipeline FPGA chips. The
interface unit consists of an Altera EPF10K20 chip and several
transceiver and buffer chips.

\subsection{Control Unit}

The control unit generate all control signals for all other units,
according to the command sent from the host computer. The control unit 
is implemented in the same FPGA chip as used for the interface unit.

\subsection{Miscellaneous Aspect}

We have packaged PROGRAPE-1 in a board of a size 37cm by 40cm.
Figure~7 is a photograph of the PROGRAPE-1 board. The total number of
chips is 84. All of the chips are wire-wrapped on the boards.  It
operates at the clock frequency of 16MHz.

\begin{figure}
\centerline{\psfig{file=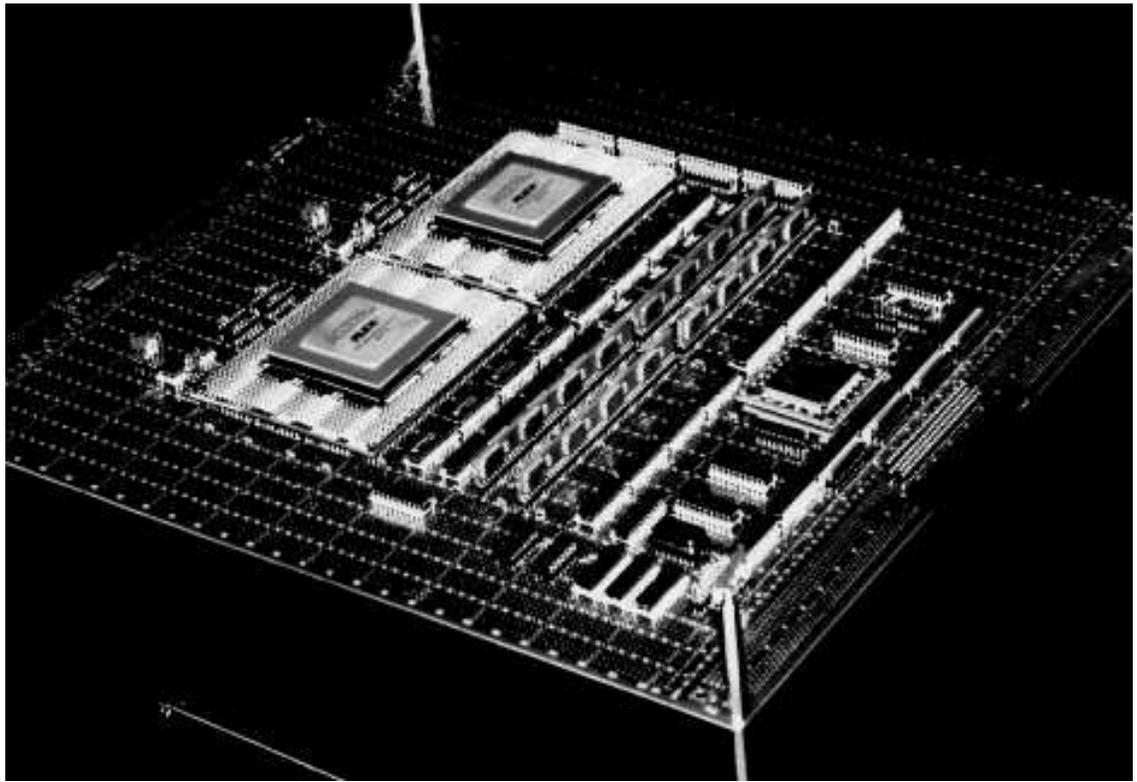,width=150mm}}
\caption{The PROGRAPE-1 board.}
\end{figure}

\section{Application of PROGRAPE-1}

In section 2.2., we discussed the general
aspect of programming a pipeline FPGA chip in PROGRAPE. Here, we discuss
practical aspects of programming, using a pipeline to calculate
gravitational interaction as an example.

\subsection{The programming model}

Figure 8 shows the internal structure of the pipeline FPGA chip. It
consists of the memory interface(MI), the I/O interface (IO), 
the pipeline unit (PU).  The PU
consists of $i$-particle register unit (IREG), the accumulator unit
(ACC), and the interaction function pipeline unit (IFP). The
application programmer need to design the IFP. All else are designed
as configurable library modules.  Both the design of IFP and the rest
of the chip are written in VHDL. Figure 9 shows the skeleton of the
top level source of the pipeline chip.

\begin{figure}
\centerline{\psfig{file=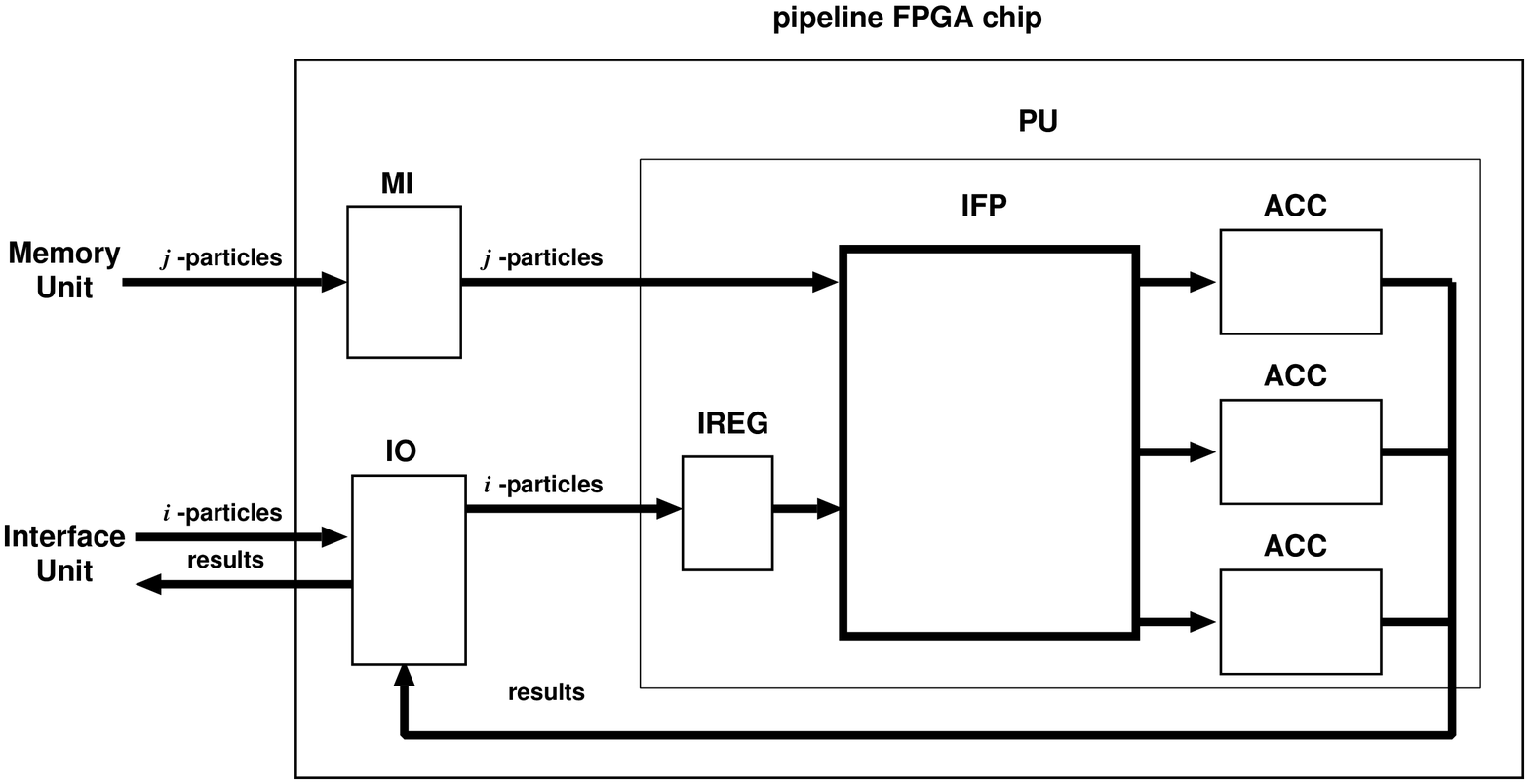,width=150mm}}
\caption{The internal structure of the pipeline chip of PRPGRAPE-1. }
\end{figure}

\begin{figure}
\begin{verbatim}
entity pipeline_top is 
    	generic ( 
        		JDATA_WIDTH : integer;
        		IDATA_WIDTH : integer;
        		FDATA_WIDTH : integer
    	);
    	port( 
    	    i_jdata : in std_logic_vector(JDATA_WIDTH-1 downto 0);
    	    i_data  : inout std_logic_vector(IDATA_WIDTH-1 downto 0);
    	    i_adr   : in std_logic_vector(9 downto 0);
    	    i_cs,i_we,i_re,i_run : in std_logic;
    	    i_clk   : in std_logic
    	);
end pipeline_top;

architecture std of pipeline_top is

signal jdata   : std_logic_vector(JDATA_WIDTH-1 downto 0);
signal datai   : std_logic_vector(IDATA_WIDTH-1 downto 0);
signal adr     : std_logic_vector(9 downto 0);
signal run, we : std_logic;
signal datao   : std_logic_vector(FDATA_WIDTH downto 0);

begin
    u1: io GENERIC MAP(FDATA_WIDTH => FDATA_WIDTH)
       PORT MAP(i_data => i_data, i_adr => i_adr,
           i_cs  => i_cs, i_we => i_we,i_re => i_re,
           i_run => i_run, clk => i_clk,
           datai => datai, adr => adr, run => run,
           we => we, datao => datao);

    u2: mi GENERIC MAP(JDATA_WIDTH => JDATA_WIDTH)
       PORT MAP(i_jdata => i_jdata, clk => i_clk, jdata => jdata);

    u3: pu GENERIC MAP(JDATA_WIDTH => JDATA_WIDTH,
        IDATA_WIDTH => IDATA_WIDTH, FDATA_WIDTH => FDATA_WIDTH)
       PORT MAP(datai => datai, jdata => jdata, we => we,
           adr => adr, run=>run, clk => i_clk,
           datao => datao);
end pipeline_top;
\end{verbatim}
\caption{The skeleton of the top level source of the pipeline chip of PROGRAPE-1. }
\end{figure}

The design flow for an application programmer would be the following

\begin{enumerate}
\item Determine the detailed specification of the pipeline, including
the number format used in each operation.
\item Verify the design using a software simulator linked to the
application program.
\item Develop the VHDL source program for the pipeline.
\item Verify the design by comparing with the simulator.
\item Integrate the VHDL source of the pipeline with other modules.
\item Generate the configuration data from the design using CAD
softwares.
\item Develop the driver software on the host computer.
\item Configure pipeline FPGA chips
\item Verify the pipeline FPGA by comparing with the application
program linked to the software simulatior.
\item Run the application program.

\end{enumerate}

We can see that a rather large amount of work is involved in
developing a pipeline FPGA chip. In principle, most of the softwares,
such as the simulator and the driver software on the host computer,
could be automatically generated from the design specification or the
VHDL source of the pipeline. We have not yet developed such software
simply because of the limitation in the available human resources.

In the rest of this section we describe  the implementation of the
gravitational interaction pipeline.

\subsection{Pipeline for Gravitational Force}

We have implemented a pipeline for the gravitational force on the
pipeline FPGA chip. It calculates the right-hand side of equation (2). 
Figure~10 shows a block diagram of the pipeline for the gravitational
force, which is almost same as the GRAPE chip \cite{omefis93}. We
omitted the circuit to handle mass of particles and the circuit to
accumulate the potential energy for simplicity.  The number formats
are the same as those used in the GRAPE chip.

\begin{figure}
\centerline{\psfig{file=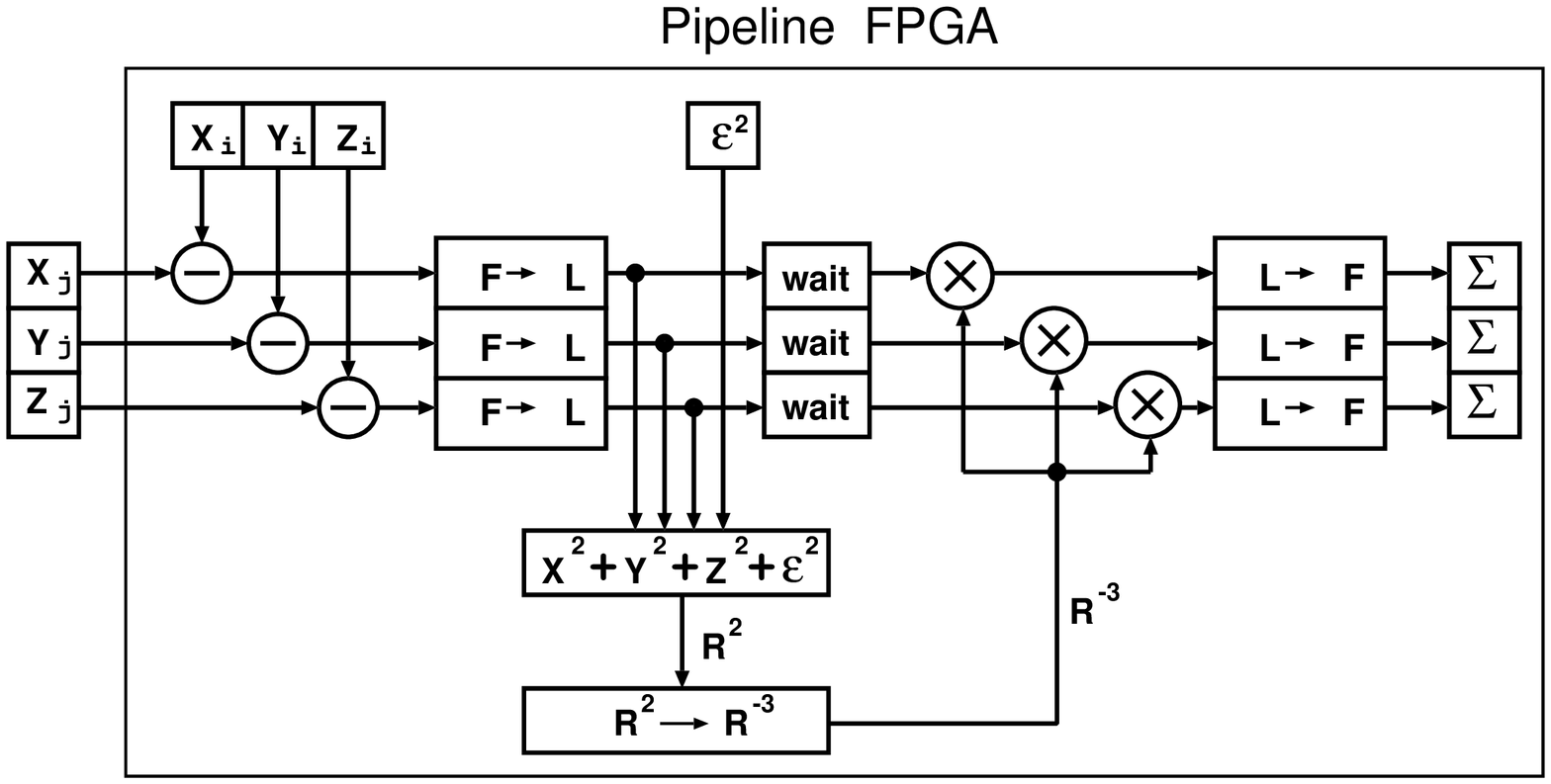,width=150mm}}
\caption{Pipeline for the gravitational force. }
\end{figure}

The pipeline for the gravitational force consumes about 50\% of the
logic cells of the pipeline FPGA chip. The nominal gate count of the
pipeline FPGA chip is 100K, while the transistor count of the GRAPE
chip was 110K. Since our implementation of the gravitational pipeline
lacks several functions implemented in the GRAPE chip, the transistor
count of the circuit equivalent to our pipeline implementation would
be around 90K. Thus, for the case of the GRAPE chip, one nominal gate
of FPGA corresponds to about 1.8 transistors. The FPGA chip
successfully operated at 16 MHz clock. Thus, assuming that the number
of floating-point operations per one interaction is 30, the peak speed 
of a chip is 480 Mflops and that of PROGRAPE-1 is 0.96 Gflops.

Whether these numbers are good or not depends on the point of
view. The FPGA chip we used contain more than one million transistors
in $0.5 {\rm \mu m}$ process, while a GRAPE chip really contains 110K
transistors in $1 {\rm \mu m}$ process. A custom LSI fabricated using
the same technology as the pipeline FPGA chip of PROGRAPE-1 would
contain around 2 pipelines operating at 100 MHz. Thus, performance
penalty of the programmability of an FPGA is as large as a factor of
100 or more. One can reduce this factor by carefully optimizing the
design of the FPGA chip and by increasing the clock period, but even
after such tuning the performance could not be three times as high as
as that of pre-optimizing design.

When compared with a general purpose computer, however, the speed
achieved with PROGRAPE-1 is pretty high. For example, a workstation
with an Alpha chip (EV56, 533MHz) has the theoretical peak speed of
1066 Mflops. However, actual speed for the calculation of the
gravitational interaction is less than 100 Mflops \cite{wgbgsw98}, even
after a heroic effort of tuning. On the other hand, PROGRAPE-1 does
deliver the performance close to the peak for real applications. In
addition, it's pretty easy to achieve massive parallelism with
PROGRAPE.


\section{Discussion}

\subsection{Comparison with other FPGA machines}

There are a number of research project to use FPGA in the form of
attached processors. As a well known and successful example, here we
discuss Splash-2 \cite{bak96}. Figure 11 shows
the architecture of a Splash-2 board. It is connected to SBus of a Sun
workstation. Up to 16 processor boards can be housed in a chassis.  The
architecture of Splash-2 is an array of basic processing units (PU). 
A PU consists of an FPGA chip and its local memory. Processing units
are connected through one shared data bus, one linear network and one
crossbar switch. Thus, we can see that Splash was designed with wide
range of applications in mind. For some applications, linear network
allowed the programmer to implement a deep pipeline which spans over
multiple FPGA chips. For other applications, the programmer can use
the 16 PUs in an SIMD fashion. It is also possible to program the
crossbar to implement a complex dataflow.

\begin{figure}
\centerline{\psfig{file=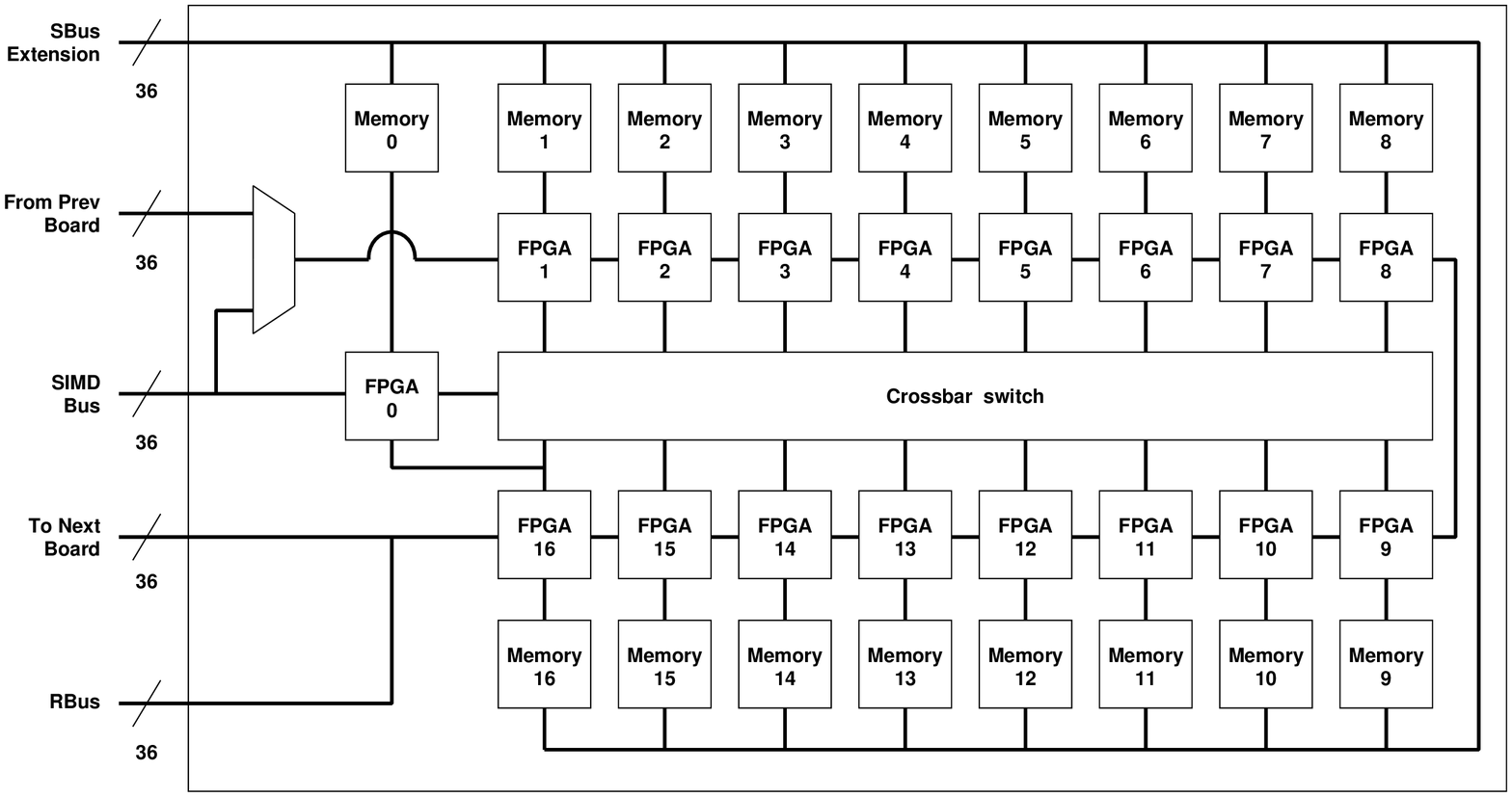,width=150mm}}
\caption{The architecture of the Splash-2 board. }
\end{figure}

With PROGRAPE, the dataflow outside the FPGA chips is fixed, and there
is only one memory unit. This design greatly simplifies the
development of the hardware. In addition, this limitation in the
programmability actually made it easier to program, provided the
architecture of PROGRAPE is suited to the target problem. Thus, as far
as the target application is the evaluation of the particle-particle
interaction, the architecture of PROGRAPE is probably better than
other more flexible architectures. Of course, more
flexible architectures can be applied to wider range of
problems. So we cannot simply say which architecture is better.

At least one group \cite{kcl95} has tried to implement the
pipeline similar to that of GRAPE-2 \cite{imfeos91} onto a
programmable FPGA board. What they used is an Altera RIPP-10 FPGA
board, which is composed of eight Altera EPF81188 FPGA chips and and
2MB of SRAM chips. They implemented all operations in 32-bit
floating-point arithmetic. A fully-parallel multiplier was too large
to be implemented, so they designed arithmetic units which require
four clock cycles to generate one result.  They implemented one
pipeline which operated at 10 MHz clock, resulting in 2.5 million
interactions per second or 75 Mflops. The nominal gate count of the
Altera RIPP-10 board was about the same as what is available on
PROGRAPE-1. The difference in the speed comes mainly from the
difference in the number format.

\subsection{Other Applications}

As we described in the introduction, our main target of PROGRAPE is
SPH calculation. Compared with the calculation of gravity, the
calculation of SPH interaction is much less expensive. Nonetheless
evaluation of the SPH interaction becomes the dominant part of the
calculation once the calculation of gravity is accelerated by GRAPE.

We are currently working on an implementation of SPH, and the result
will be described elsewhere. Here, we discuss what else would be
suitable for PROGRAPE.

Short-range particle-particle interactions are good candidate for the
implementation on PROGRAPE. Those include, beside SPH, van-der-Waals
force in molecular dynamics calculation and Coulomb force with cutoff
used in Ewald method or P$^3$M method. As we have seen, the actual
performance of an FPGA depends very strongly on the required
accuracy. Since relatively little is known about what level of the
accuracy is required for these calculation, we cannot predict whether
PROGRAPE is useful for these application or not.

Though the overall architecture of PROGRAPE is determined so that it
is optimized for the evaluation of particle-particle interaction, it
can be used for other operations which can be expressed in the form of
equation (1). One example is the discrete Fourier
transform used in the direct Ewald method. Both the discrete Fourier
transform and the inverse transform can be expressed in the form of
equation (1) \cite{fmioes93}, and therefore can be efficiently
implemented on PROGRAPE. In the case of the Ewald method, the required 
accuracy is relatively low (since it handles  higher-order
correction of the force due to  ``image'' particles). Thus PROGRAPE would 
be a good alternative to an implementation on a general-purpose
computer.

It is also possible to use PROGRAPE hardware for applications which
cannot be expressed in the form of equation (1). As
one example, consider the matrix multiplication
\begin{equation}
C = A\cdot B
\end{equation}
We can use the memory unit to store matrix $A$, and internal memory
available in FPGA chips to store one row of matrix $B$. Thus, we can
perform the multiplication in the so-called inner-product form. We can 
also use the same configuration to implement the Gaussian
elimination. Here again, the consideration of the accuracy would
determine the effectiveness of PROGRAPE.

\subsection{Future Prospect}

We plan to develop a massively parallel PROGRAPE, the PROGRAPE-2. It
will consist of around 1000 FPGA chips, and 2-4 FPGA chips will share
the memory unit. The overall architecture will be the same as that of
GRAPE-6 \cite{m99}. The only difference between GRAPE-6 and
PROGRAPE-2 will be that the FPGA chips are used in the place of the
GRAPE-6 chip. The performance of PROGRAPE-2 system will vary depending 
on applications, but it will be in the range of 1-10 Tflops.  We plan
to complete PROGRAPE-2 by 2001.

\end{document}